\documentclass[journal=jpccck,manuscript=article]{achemso}

\usepackage[english]{babel}
\usepackage[utf8]{inputenc}
\usepackage{amsmath}
\usepackage{booktabs}
\usepackage{enumitem}
\usepackage{tikz}
\usepackage{subcaption}
\usetikzlibrary{positioning,arrows.meta}

\usepackage{hyperref}
\hypersetup{
  colorlinks=true,
  citecolor=blue,
  linkcolor=blue,
  urlcolor=blue
}

\graphicspath{{../analysis/}{analysis/}{./}}

\title{Interpreting Latent Electrostatics in Charge-Aware Machine Learning Potentials for Battery Electrolytes}

\author{Abibat Adekoya-Olowofela}
\affiliation[UIC]{Department of Mechanical and Industrial Engineering,
  University of Illinois Chicago, Chicago, Illinois 60607, United States}
\alsoaffiliation[CNM]{Center for Nanoscale Materials,
  Argonne National Laboratory, Lemont, Illinois 60439, United States}

\author{Sukriti Manna}
\affiliation[UIC]{Department of Mechanical and Industrial Engineering,
  University of Illinois Chicago, Chicago, Illinois 60607, United States}
\alsoaffiliation[CNM]{Center for Nanoscale Materials,
  Argonne National Laboratory, Lemont, Illinois 60439, United States}

\author{Shola Folarin}
\affiliation[UIC]{Department of Mechanical and Industrial Engineering,
  University of Illinois Chicago, Chicago, Illinois 60607, United States}
\alsoaffiliation[CNM]{Center for Nanoscale Materials,
  Argonne National Laboratory, Lemont, Illinois 60439, United States}

\author{Mikhail Trought}
\affiliation[Ford]{Ford Motor Company, Dearborn, Michigan 48124, United States}

\author{Benjamin Emley}
\affiliation[Ford]{Ford Motor Company, Dearborn, Michigan 48124, United States}

\author{Ann Straccia}
\affiliation[Ford]{Ford Motor Company, Dearborn, Michigan 48124, United States}

\author{Avinesh Ojha}
\email{aojha@ford.com}
\affiliation[Ford]{Ford Motor Company, Dearborn, Michigan 48124, United States} 

\author{Subramanian K.R.S. Sankaranarayanan}
\email{skrssank@uic.edu}
\affiliation[UIC]{Department of Mechanical and Industrial Engineering,
  University of Illinois Chicago, Chicago, Illinois 60607, United States}
\alsoaffiliation[CNM]{Center for Nanoscale Materials,
  Argonne National Laboratory, Lemont, Illinois 60439, United States}

\author{Xueping Qin}
\email{xqin@anl.gov}
\affiliation[CNM]{Center for Nanoscale Materials,
  Argonne National Laboratory, Lemont, Illinois 60439, United States}

\begin{document}

\begin{abstract}

Long-range electrostatic interactions strongly influence the structure,
dynamics, and transport of liquid electrolytes. Although charge-aware machine-learned
interatomic potentials (MLIPs) enable a more explicit treatment of
these interactions, the physical meaning of their learned
electrostatic variables remains unclear. Here, we develop and analyze the latent
electrostatic representation learned by a Moment Tensor Potential with Latent
Electrostatics (MTP--LES) for a LiPF$_6$/EC/EMC electrolyte. The model is
trained exclusively on density functional theory (DFT) energies and forces, without
supervision on atomic charges, dipole moments, Born effective charges, or
ion-pair labels. We find statistically robust, practically large, and structure-resolved differences between
contact ion-pair and solvent-separated ion-pair environments in the latent Li charge,
the coordinating-oxygen response, and the long-range LES energy, with effect sizes
exceeding $|d|=1$ and 95\% bootstrap confidence intervals
that exclude zero. Within the model's energy partition, the associated ensemble has an LES contribution approximately $0.5$~eV lower and a short-range contribution approximately $0.9$~eV higher than the dissociated ensemble, revealing compensating long- and short-range contributions that are obscured in the total energy alone. The coordinating-oxygen response remains distinct after accounting for
O--Li distance, indicating sensitivity to the broader ion-pairing and solvation environment 
rather than to nearest-neighbor geometry alone. 
These findings establish that latent electrostatic variables learned solely from energies and forces can distinguish ion-pairing and solvation environments in a liquid electrolyte, providing a basis for evaluating the learned electrostatic representation of charge-aware MLIPs against independent structural observables.

\end{abstract}

%%%%%%%%%%%%%%%%%%%%%%%%%%%%%%%%%%%%%%%%%%%%%%%%%%%%%%%%%%%%%%%%%%%%%%%%%%%%%%%
%%%%%%%%%%%%%%%%%%%%%%%%%%%%%%%%%%%%%%%%%%%%%%%%%%%%%%%%%%%%%%%%%%%%%%%%%%%%%%%
\section{Introduction}
%%%%%%%%%%%%%%%%%%%%%%%%%%%%%%%%%%%%%%%%%%%%%%%%%%%%%%%%%%%%%%%%%%%%%%%%%%%%%%%

Electrolytes play a central role in electrochemical energy-storage
technologies, including lithium-ion batteries, where ion transport, solvation,
and ion association influence ionic conductivity, interfacial stability, and
electrochemical performance~\cite{Xu2004}. Capturing these phenomena at the
atomistic level requires an accurate description of both short-range chemical
interactions and long-range electrostatic forces. Density functional theory
(DFT) provides a first-principles description of these interactions, but its
computational cost restricts simulations to relatively small systems and short
timescales. Machine-learned interatomic potentials (MLIPs) offer a route to
larger-scale molecular dynamics while retaining near first-principles
accuracy~\cite{Deringer2019}. Early approaches, including the Moment Tensor
Potential (MTP) and high-dimensional neural network potentials, demonstrated
that representations of local atomic environments can accurately reproduce
quantum-mechanical potential-energy surfaces~\cite{Shapeev2016,Behler2007}.
More recent graph-based and equivariant architectures, including
NequIP~\cite{Batzner2022}, MACE~\cite{NEURIPS2022_4a36c3c5}, and
CACE~\cite{Cheng2024}, have further advanced the accuracy and transferability
of MLIPs across diverse classes of materials. Long-range electrostatic
interactions remain challenging because they extend beyond the
finite local environments used by most MLIPs. This limitation is particularly
important in systems where polarization, dielectric response, ion association,
and solvation are strongly influenced by electrostatics.

To address this limitation, charge-aware MLIPs incorporate electrostatic
interactions through fixed or environment-dependent atomic charges,
charge-equilibration schemes, or latent electrostatic representations.
Examples include fourth-generation high-dimensional neural network potentials
(4G-HDNNPs)~\cite{Ko2021}, PhysNet~\cite{Unke2019}, qNEP~\cite{Fan2026}, and
the QRd and EDQRd extensions of the Moment Tensor
Potential~\cite{Korogod2026,KorogodEDQRd2026}. A complementary approach is
Latent Ewald Summation (LES)~\cite{Cheng2025LES,Zhong2025}, in which
environment-dependent latent charges are inferred from local atomic
representations and coupled through an Ewald summation while requiring
supervision only from reference energies and forces. LES has been formulated
as a general augmentation layer and combined with several short-range
architectures, including CACE, MACE, NequIP, Allegro, and
CHGNet~\cite{Kim2025LESaug}. Our recently developed Moment Tensor Potential
with Latent Electrostatics (MTP--LES)~\cite{AdekoyaOlowofela2026MTPLES}
extends this strategy to invariant moment tensor descriptors.

Recent work by Chalykh \emph{et al.} examined explicit electrostatics within
MTP-based models for EC/EMC mixtures and LiPF$_6$/EC/EMC electrolytes,
comparing conventional MTP with fixed-charge MTP--QRd and
environment-dependent MTP--EDQRd formulations~\cite{Chalykh2026}. Their study
examined active-learning dataset construction, predictive accuracy,
molecular-dynamics stability, transferability, electrolyte structure, and
ionic conductivity, and showed advantages of environment-dependent charges
over the fixed-charge formulation for force accuracy and simulation stability.
The learned charge variables themselves were not analyzed as
structure-resolved descriptors of ion association or local solvation. This
leaves a complementary question: do electrostatic variables learned solely
from energy and force supervision retain interpretable information about
the environments from which they emerge.

This question is important because latent charges are not uniquely defined
quantum-mechanical observables. Even conventional atomic partial charges
depend on how the electronic density is partitioned, with methods such as
Bader, Hirshfeld, and DDEC assigning different atomic charges to the same
underlying electronic structure~\cite{Bader1985,Hirshfeld1977,Manz2016}.
Latent charges learned without direct charge supervision therefore should not
automatically be identified with conventional physical partial charges.
Their usefulness as internal descriptors instead depends on whether their
variations show reproducible relationships with independently defined
structural environments.

Liquid electrolytes provide a useful setting in which to examine this
relationship. Contact ion pairs (CIPs) and solvent-separated ion pairs
(SSIPs) differ in Li$^+$--anion separation and in the surrounding solvent
coordination. These structural changes provide independent reference
descriptors against which a learned electrostatic response can be evaluated.
If latent electrostatic variables respond reproducibly to these environments,
they could provide information about ion association and solvation without
requiring additional electronic-structure calculations or post hoc
charge-partitioning analyses. In our recent work, we introduced MTP--LES as a framework for incorporating
long-range electrostatics into invariant machine-learned interatomic
potentials~\cite{AdekoyaOlowofela2026MTPLES}. We showed that the model
reproduces reference energies and forces and that its latent electrostatic
representation supports the prediction of electrical-response and structural
properties, including Born effective charges, dielectric response, and
infrared spectra, across chemically diverse systems. Those results established
the predictive capability of the framework but did not address what
information the latent electrostatic variables themselves retain about
different local chemical environments.

To our knowledge, it has not been systematically
examined whether electrostatic variables learned
solely from energies and forces retain structure-resolved information about ion
association and solvation in battery electrolytes. Here, we investigate this question using a LiPF$_6$/EC/EMC
electrolyte dataset containing associated contact ion-pair and dissociated
solvent-separated ion-pair configurations. The MTP--LES model is trained only
on DFT energies and forces, without supervision on atomic charges,
electrostatic potentials, dipole moments, coordination numbers, or ion-pair
labels. We examine the learned electrostatic representation alongside
independently computed structural descriptors, including Li--P separation,
Li--O coordination, and local solvent geometry, and use configuration-level
statistical analyses to quantify differences between the two ion-pair
environments.

We find that the associated and dissociated ensembles exhibit robust
differences in the latent electrostatic response of Li$^+$, PF$_6^-$, and
coordinating solvent oxygen atoms. Decomposition of the model energy further
reveals opposing short- and long-range contributions that are partly obscured
in the total energy, while the coordinating-oxygen response remains distinct
after accounting for nearest-neighbor O--Li geometry. These results show that
electrostatic variables learned solely through energy and force supervision
retain structure-resolved signatures of ion association and solvation. Rather
than assigning these variables the status of uniquely defined physical partial
charges, we evaluate their meaning through reproducible relationships with
independent structural quantities.

\section{Results}

We first establish the predictive accuracy of MTP--LES on the
LiPF$_6$/EC/EMC electrolyte dataset and then ask whether the learned
electrostatic representation captures structure-resolved signatures of ion
association and local solvation. Unless otherwise stated, all analyses are
performed on the held-out test set. The Results proceed in four stages.
First, we introduce the two ion-pair environments sampled by the dataset and
show that the trained model reproduces the DFT reference energies and forces.
Second, we characterize the structural differences between the associated and
dissociated ensembles using descriptors that are independent of the model.
Third, we show that the latent charges and the decomposition of the total energy
into short- and long-range contributions carry systematic, ensemble-dependent
electrostatic signatures.
Fourth, we demonstrate that this response extends
to the coordinating solvent and quantify all ensemble-level differences
statistically. Together, these analyses show that a model supervised only on
energies and forces develops an internal electrostatic representation that
tracks the ion-pairing state and its solvation environment.

%%%%%%%%%%%%%%%%%%%%%%%%%%%%%%%%%%%%%%%%%%%%%%%%%%%%%%%%%%%%%%%%%%%%%%%%%%%%%%%
\subsection{Model system and validation}
%%%%%%%%%%%%%%%%%%%%%%%%%%%%%%%%%%%%%%%%%%%%%%%%%%%%%%%%%%%%%%%%%%%%%%%%%%%%%%%

\begin{figure}[H]
    \centering
    \includegraphics[width=\linewidth]{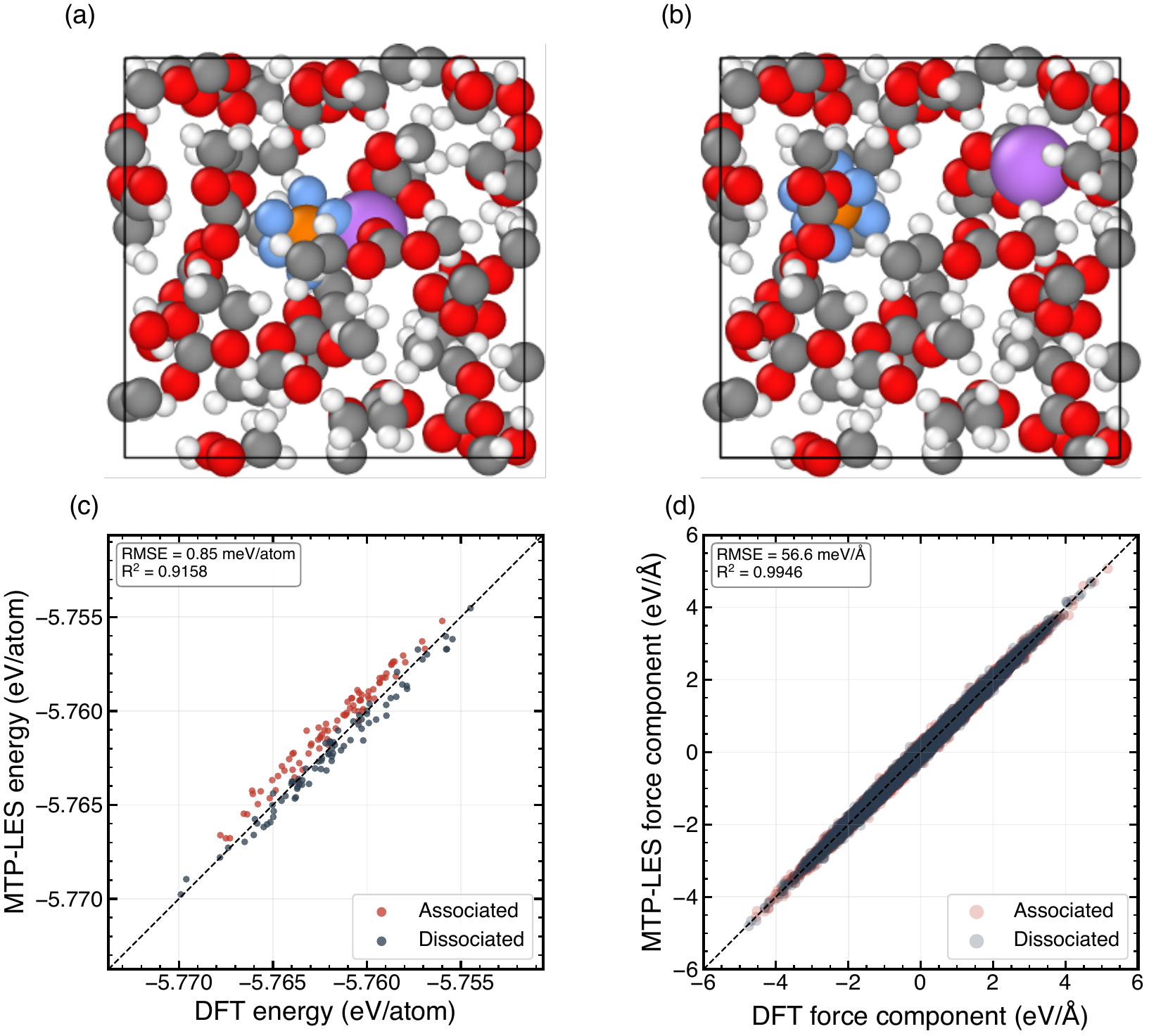}

    \caption{\textbf{Model system and validation of the MTP--LES model.}
    (a)~Contact ion pair (CIP, associated), in which Li$^+$ remains in direct
    contact with PF$_6^-$. 
    (b)~Solvent-separated ion pair (SSIP, dissociated), in which one or more
    solvent molecules separate Li$^+$ from PF$_6^-$, resulting in a more
    strongly solvated Li$^+$ environment. 
    (c)~Parity plot comparing MTP--LES predictions with DFT reference energies
    per atom on the held-out test set. 
    (d)~Parity plot comparing MTP--LES predictions with DFT reference force
    components on the held-out test set. Both associated and dissociated configurations are represented in the
held-out test set. Color code: Li, purple; P, orange; F, blue; O, red; C, grey; H, white.}

    \label{fig:validation}
\end{figure}

The LiPF$_6$/EC/EMC dataset samples two dominant ion-association
environments: contact ion pairs (CIPs), in which Li$^+$ remains in direct
contact with the PF$_6^-$ anion while also interacting with
surrounding solvent molecules, and solvent-separated ion pairs (SSIPs), in
which one or more solvent molecules separate Li$^+$ from PF$_6^-$ and the
cation is more fully solvated. Representative configurations from the two
ensembles are shown in Figure~\ref{fig:validation}a,b. Throughout this work,
we refer to the CIP ensemble as associated and the SSIP ensemble as
dissociated; these two local environments provide the structural basis for
all analyses that follow.

The MTP--LES model was trained on DFT energies and forces of configurations
drawn from both ensembles, and the checkpoint with the lowest validation
error was retained (Methods). Before interpreting the learned electrostatic
representation, we assess the predictive accuracy of this model on the
held-out test set. Figure~\ref{fig:validation}c,d compares the predicted and
reference DFT energies and force components. MTP--LES reproduces the DFT
potential energy surface with an energy RMSE of approximately
0.85~meV\,atom$^{-1}$ and a force RMSE of 56.6~meV\,\AA$^{-1}$. Errors for the associated and dissociated configurations, including force
errors by element, are given in Supplementary Table~S1. This predictive accuracy provides a basis for examining whether the model's learned electrostatic variables encode signatures of ion association and solvation.

%%%%%%%%%%%%%%%%%%%%%%%%%%%%%%%%%%%%%%%%%%%%%%%%%%%%%%%%%%%%%%%%%%%%%%%%%%%%%%%
\subsection{Structural characterization of ion-pair ensembles}
%%%%%%%%%%%%%%%%%%%%%%%%%%%%%%%%%%%%%%%%%%%%%%%%%%%%%%%%%%%%%%%%%%%%%%%%%%%%%%%

\begin{figure}[H]
\centering
\includegraphics[width=\linewidth]{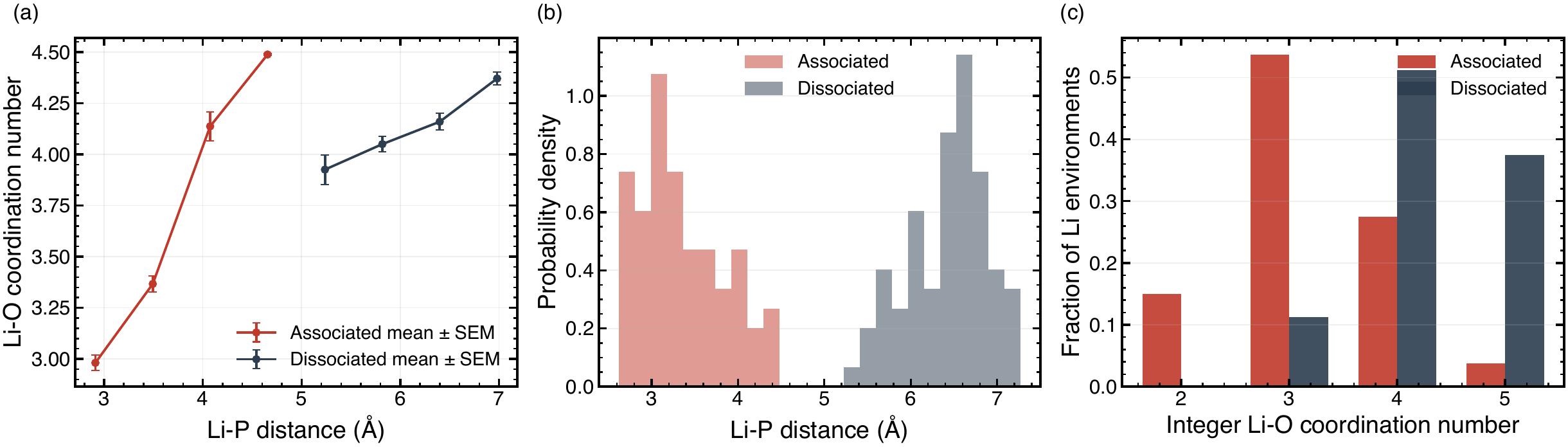}
\caption{\textbf{Structural characterization of ion association.}
(a)~Continuous Li--O coordination number as a function of Li--P separation,
shown as state-specific binned means $\pm$~SEM.
(b)~Probability distributions of the Li--P separation for the associated and
dissociated ensembles.
(c)~Distribution of integer Li--O coordination numbers. Panel (a) uses a
continuous coordination measure, whereas panel (c) and
Table~\ref{tab:statistics} use integer coordination numbers, leading to
small differences in their absolute values. Together, these descriptors
characterize the distinct ion-pair and solvation environments sampled by
two ensembles.}
\label{fig:coordination}
\end{figure}

The structural differences between the two ion-pair environments are
quantified in Figure~\ref{fig:coordination}. As expected, the associated and
dissociated ensembles occupy distinct Li--P distance ranges, consistent with
their respective ion-association states
(Figure~\ref{fig:coordination}a,b). Dissociated configurations also exhibit
larger Li--O coordination numbers on average, because the departure of the
anion from the first shell leaves room for additional solvent molecules
around Li$^+$ (Figure~\ref{fig:coordination}a,c). These ensemble-level
differences confirm two distinct ion-pair and solvation environments.
Because the ensembles were generated from separate trajectories, they should not imply a
continuous structural pathway between the two states. Importantly, these
structural descriptors are computed from geometry alone: they are
independent of the MTP--LES model and are not used during training. They
therefore provide reference measurements against which the learned
electrostatic response can be evaluated.

%%%%%%%%%%%%%%%%%%%%%%%%%%%%%%%%%%%%%%%%%%%%%%%%%%%%%%%%%%%%%%%%%%%%%%%%%%%%%%%
\subsection{Latent electrostatic signatures emerge from energy and force
supervision}
%%%%%%%%%%%%%%%%%%%%%%%%%%%%%%%%%%%%%%%%%%%%%%%%%%%%%%%%%%%%%%%%%%%%%%%%%%%%%%%

\begin{figure}[H]
\centering
\includegraphics[width=\linewidth]{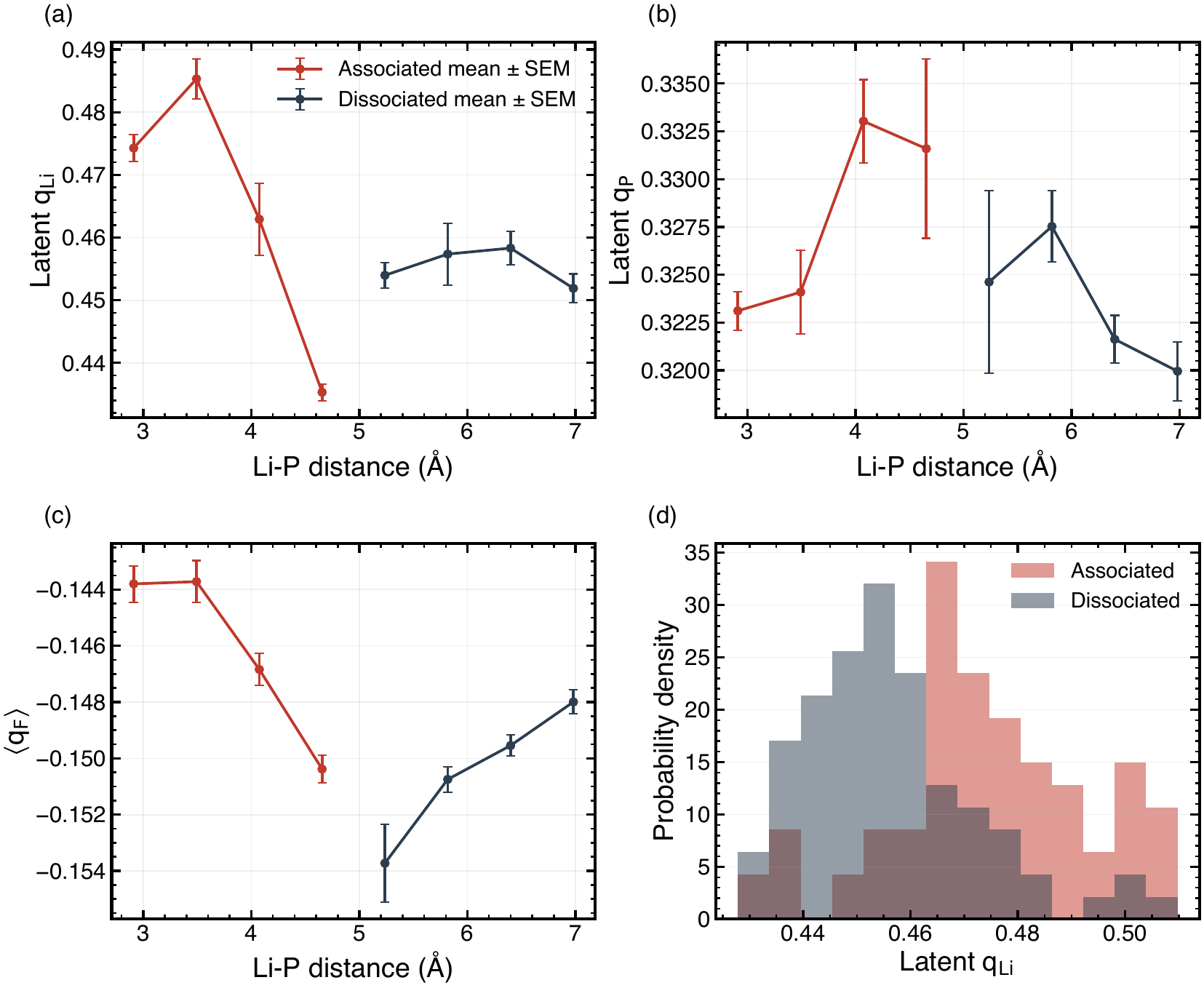}
\caption{\textbf{Latent electrostatic signatures of ion association.}
Latent (a)~Li, (b)~P, and (c)~mean F charges as functions of Li--P
separation, shown as state-specific binned means with error bars denoting
the standard error of the mean (SEM). The associated and dissociated
ensembles were binned separately, resulting in two disconnected curves.
(d)~Distribution of latent Li charges for the two ensembles. Associated
configurations exhibit a larger latent Li charge on average, although the
distributions overlap. The ensemble-dependent shifts across Li and F emerge despite the absence of direct supervision on atomic charges.}
\label{fig:charges}
\end{figure}

Having established the structural differences between the associated and
dissociated environments, we next examine whether the learned electrostatic
response reflects these changes despite receiving no direct electrostatic
supervision during training. Figure~\ref{fig:charges}a shows the latent Li
charge as a function of Li--P separation, and Figure~\ref{fig:charges}d shows the
corresponding charge distributions for these two ensembles. Associated
configurations exhibit larger latent Li charges on average than dissociated
configurations, although the two distributions overlap substantially. The
mean shift between the ensembles is approximately $0.018$ on the internal
latent-charge scale and is highly reproducible across the held-out test
configurations (Table~\ref{tab:statistics}).

The dependence of the latent Li charge on Li--P separation within each
ensemble is more complex (Figure~\ref{fig:charges}a). Within the associated
branch, the latent Li charge initially increases and then decreases toward
the upper end of the sampled Li--P range, whereas the dissociated branch
varies only weakly. The robust result is therefore the ensemble-level 
difference itself, rather than any simple monotonic dependence of
the latent charge on Li--P separation.

The ensemble-dependent response is not confined to the cation. The mean F
charge of the anion also differs between the two ensembles, and the latent P charge varies with Li--P separation (Figure~\ref{fig:charges}b,c), showing
that the electrostatic response to ion pairing extends across the PF$_6^-$
anion. Because
overall charge neutrality is enforced by construction, the latent atomic
charges are not mutually independent, and their correlated variation should
not be interpreted as directional charge transfer between particular species. The detailed turning points of the P and F responses near the boundary between
the two ensembles also lie in sparsely populated bins ($n=4$ and $n=3$
configurations) and should not be assigned mechanistic significance. Since no atomic charges, electrostatic
potentials, dipole moments, Born effective charges, or ion-pair labels were
supplied as training targets, these ensemble-dependent shifts emerge
indirectly from optimization of DFT energies and forces, indicating that the
learned representation encodes information associated with distinct local
ion-pair environments.

%%%%%%%%%%%%%%%%%%%%%%%%%%%%%%%%%%%%%%%%%%%%%%%%%%%%%%%%%%%%%%%%%%%%%%%%%%%%%%%
\subsection{Latent electrostatic energy reveals short- and long-range
compensation}
%%%%%%%%%%%%%%%%%%%%%%%%%%%%%%%%%%%%%%%%%%%%%%%%%%%%%%%%%%%%%%%%%%%%%%%%%%%%%%%

\begin{figure}[H]
\centering
\includegraphics[width=\linewidth]{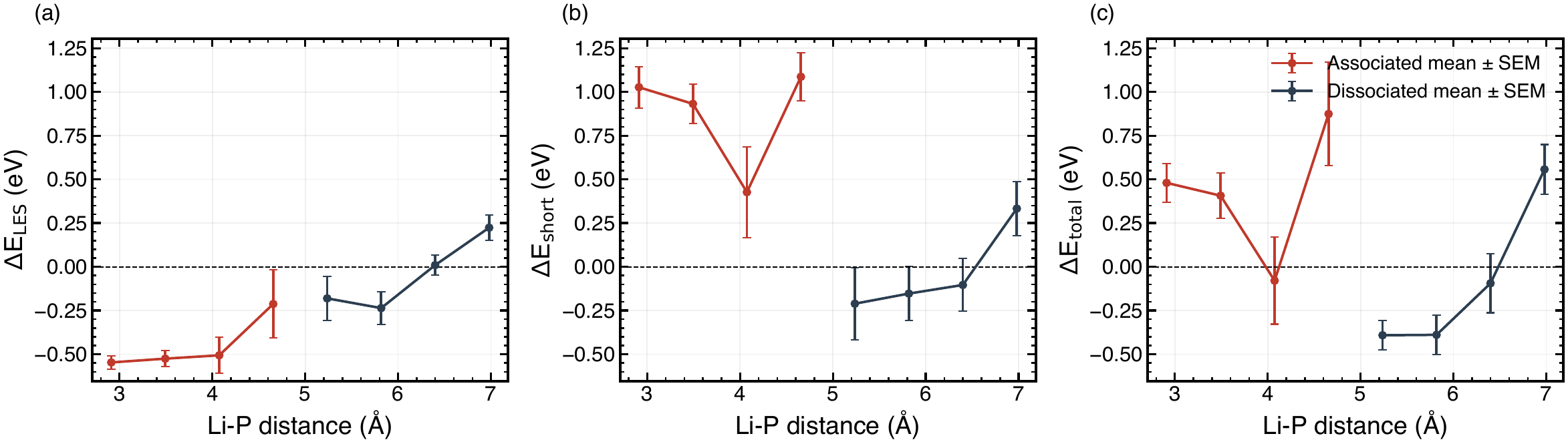}
\caption{\textbf{Energy decompositions across the sampled ion-association
environments.}
(a)~Latent LES electrostatic energy, (b)~short-range energy contribution,
and (c)~total energy as functions of Li--P separation for the associated and
dissociated ion-pair ensembles. Energies are referenced to the mean energy
of the dissociated ensemble. Associated configurations exhibit more favorable LES contributions on average
but higher short-range energies, so that the two contributions partially
compensate within the model's energy partition. Error bars denote the standard error of the mean
(SEM).}
\label{fig:stabilization}
\end{figure}

The learned representation provides access not only to the latent atomic
charges but also to the long-range LES contribution to the total energy.
Figure~\ref{fig:stabilization} shows the LES, short-range, and total energy
contributions as a function of Li--P separation. Associated configurations
have lower LES energies on average than the dissociated configurations
(Figure~\ref{fig:stabilization}a), with a mean difference of approximately
$0.52$~eV between the two ensembles (Table~\ref{tab:statistics}). This
difference includes both the reciprocal-space sum and the self term of
Eq.~\eqref{eq:les}. The short-range contribution shows
the opposite behavior, with higher energies for the associated configurations
(Figure~\ref{fig:stabilization}b). The variation within each branch is not
strictly monotonic. The bins nearest the crossover are sparsely populated
($n=4$ and $n=3$) and therefore carry
larger uncertainties, so their local variation should not be overinterpreted.

Within the model, the two energy contributions partially compensate for one
another. Relative to the dissociated ensemble, the associated ensemble has an
LES contribution lower by 0.517~eV and a short-range contribution higher by
0.915~eV. The model's energy partition therefore assigns the associated ensemble a more
favorable long-range electrostatic contribution and a less favorable
short-range contribution on average, a distinction that is not visible in the
total energy alone. The lower LES contribution provides an additional
signature of ion association within the learned representation. The predicted mean total energies of the associated and dissociated ensembles
($-1601.50$ and $-1601.90$~eV per cell) differ from the DFT reference values
($-1601.77$ and $-1601.82$~eV) by less than 1~meV\,atom$^{-1}$.
Because the reference calculations provide no such decomposition, we interpret
these contrasts as properties of the model's internal partition.

The associated and dissociated configurations were sampled as separate
ensembles, with only sparse coverage of the Li--P distances separating the
two states. The energy differences reported here should therefore not be
interpreted as a continuous dissociation pathway, a thermodynamic preference,
or a free-energy profile. Establishing any of these would require denser
sampling across the transition region together with the corresponding
entropic contributions.

%%%%%%%%%%%%%%%%%%%%%%%%%%%%%%%%%%%%%%%%%%%%%%%%%%%%%%%%%%%%%%%%%%%%%%%%%%%%%%%
\subsection{Solvent coordination and the coordinating-oxygen response}
%%%%%%%%%%%%%%%%%%%%%%%%%%%%%%%%%%%%%%%%%%%%%%%%%%%%%%%%%%%%%%%%%%%%%%%%%%%%%%%

\begin{figure}[H]
\centering
\includegraphics[width=\linewidth]{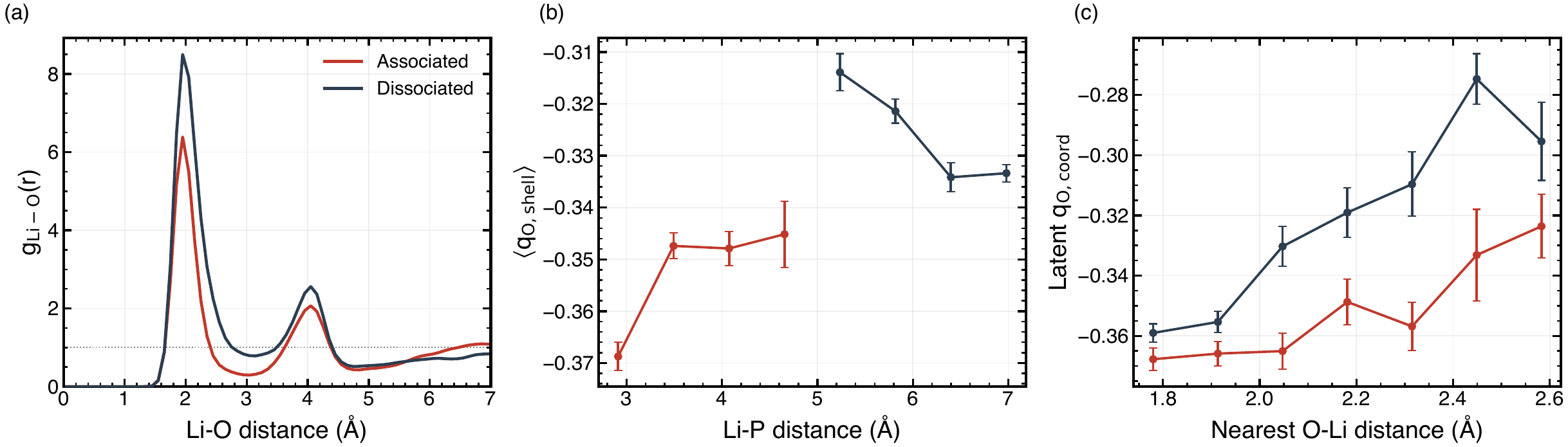}
\caption{\textbf{Solvent structure and coordinating-oxygen latent response.}
(a)~Li--O radial distribution functions for the associated and dissociated
ensembles.
(b)~Mean latent charge per first-shell oxygen atom as a function of Li--P
separation, shown as state-specific binned means $\pm$~SEM.
(c)~Latent charge of first-shell oxygen atoms as a function of the nearest
O--Li distance, shown as state-specific binned means $\pm$~SEM.}
\label{fig:solvation}
\end{figure}

We next ask how the learned electrostatic response relates to the local
solvent environment. Figure~\ref{fig:solvation}a shows the Li--O radial
distribution function for the two ensembles. Both ensembles show the first
Li--O peak at approximately 1.95~\AA{}, indicating that the preferred Li--O
contact distance is essentially unchanged between the associated and
dissociated states. The weaker peak in the associated ensemble is consistent
with partial displacement of solvent from the Li$^+$ coordination shell by
the nearby anion. Together with the coordination statistics of
Figure~\ref{fig:coordination}, this supports the structural picture that
dissociated Li$^+$ is coordinated by more solvent oxygen atoms, consistent
with solvent replacing the anion in the first shell upon dissociation.

The latent response of the coordinating solvent tracks this environment, but
not through coordination number alone. The mean latent charge per
first-shell oxygen remains more negative in the associated ensemble
throughout its sampled Li--P range, whereas the dissociated ensemble occupies
a distinctly less negative range (Figure~\ref{fig:solvation}b). Because this
quantity is normalized per coordinating oxygen, it is not simply a restatement of
the coordination number: although the two ensembles overlap in Li--O
coordination, their ranges of mean first-shell oxygen charge remain
separated.

The same conclusion follows from a local analysis.
Figure~\ref{fig:solvation}c shows the latent charge of first-shell oxygen
atoms as a function of the nearest O--Li separation. The latent oxygen
charge generally becomes less negative as the nearest O--Li distance
increases, although the dependence is not strictly monotonic. More
importantly, the two ensembles do not collapse onto a single common curve: the
associated ensemble exhibits a more negative latent oxygen response across
the overlapping O--Li distance range, with the offset becoming smaller at
the shortest separations. After accounting for the O--Li distance
distribution, the associated-minus-dissociated difference in
coordinating-oxygen latent charge is $-0.028$ [$-0.034$, $-0.022$];
restricting the analysis to oxygens beyond $2.0$~\AA{} and averaging within
each configuration gives $-0.037$ [$-0.048$, $-0.026$]. The two ensembles
have nearly identical mean O--Li distances (2.13 and 2.10~\AA{}), so
nearest-neighbor geometry alone cannot account for the offset. As a further control, the difference between the mean latent charges of
coordinating and non-coordinating oxygen atoms within each configuration
gives an associated-minus-dissociated contrast of $-0.027$
[$-0.033$, $-0.022$] (Cohen's $d=-1.61$). The uniform shift applied by the
neutrality projection (Eq.~\eqref{eq:neutrality}) cancels in this
within-configuration difference, so the oxygen response cannot be attributed
to that shift alone. The pooled coordinating-oxygen contrast also reflects
differences in the proportions of coordinating oxygen types
(Supplementary Table~S2).

Two features of the data limit how far these observations can be pushed.
The highest-separation associated bin and the lowest-separation dissociated
bin contain only four and three configurations, respectively, and their local
variation should therefore not be overinterpreted. In addition, because
charge neutrality is enforced by construction, the latent charges are coupled
across the system. Their variation reflects the learned electrostatic response
to changes in the local environment, but the individual values should not be
interpreted as uniquely defined physical partial charges. The key result 
is therefore the systematic difference between the two ensembles across
several model-derived electrostatic quantities, evaluated alongside
independently computed structural descriptors. Taken together with the anion
response in Figure~\ref{fig:charges}, these results show that MTP--LES encodes
the ion-pair state through a collective electrostatic response involving the
cation, anion, and coordinating solvent, rather than through the latent Li
charge alone or any single local structural descriptor.

%%%%%%%%%%%%%%%%%%%%%%%%%%%%%%%%%%%%%%%%%%%%%%%%%%%%%%%%%%%%%%%%%%%%%%%%%%%%%%%
\subsection{Statistical comparison of ion-pair ensembles}
%%%%%%%%%%%%%%%%%%%%%%%%%%%%%%%%%%%%%%%%%%%%%%%%%%%%%%%%%%%%%%%%%%%%%%%%%%%%%%%

The structural and electrostatic differences between the associated and
dissociated environments were quantified using configuration-level bootstrap
resampling and Cohen's $d$ effect sizes (Table~\ref{tab:statistics}). The
two ensembles are structurally distinct: associated configurations have
shorter Li--P separations, whereas dissociated configurations have larger
Li--O coordination numbers, consistent with increased solvent coordination
upon dissociation. The very large effect size for the Li--P separation
(Cohen's $d = -6.39$) should be interpreted as a consistency check rather than a
discovered effect, since the ensembles originate from trajectories sampling
distinct ion-pairing states and this coordinate differs between them by
construction.

The learned electrostatic quantities exhibit equally robust ensemble-level
differences. Associated configurations have larger latent Li charges, more
negative coordinating-oxygen latent charges, and more favorable LES energies
than dissociated configurations. The shift in latent Li charge is modest in
magnitude but systematic across the held-out test set, giving a large effect
size (Cohen's $d = +1.04$) despite substantial overlap between the two
distributions. All reported differences have 95\% confidence intervals that
exclude zero, and for all quantities except the total model energy, no
bootstrap replicate crossed zero among 10{,}000 resamples.

\begin{table}[H]
\centering
\caption{\textbf{Statistical comparison of the associated and dissociated
ensembles.}
Values are reported as mean $\pm$ standard deviation. Differences are defined
as associated minus dissociated, with 95\% bootstrap confidence intervals.
Cohen's $d$ gives the standardized effect size. The Li--P separation differs
between the ensembles by
construction and is included as a consistency check. Latent charges are
reported on the internal MTP--LES scale and are not interpreted as calibrated
physical partial charges. $E_{\mathrm{short}}$ denotes the short-range
contribution with the constant training-energy reference restored, such that
$E_{\mathrm{LES}} + E_{\mathrm{short}} = E_{\mathrm{total}}$. Model
energies are reported per simulation cell. Statistical procedures are
described in the Methods.}
\label{tab:statistics}

\footnotesize
\setlength{\tabcolsep}{6pt}
\renewcommand{\arraystretch}{1.2}

\begin{tabular}{lcccc}
\toprule
\textbf{Quantity} &
\textbf{Associated} &
\textbf{Dissociated} &
\textbf{$\Delta$ (95\% CI)} &
\textbf{Cohen's $d$} \\
\midrule

$\langle r_{\mathrm{Li-P}}\rangle$ (\AA)
& $3.354 \pm 0.496$
& $6.419 \pm 0.463$
& $-3.065$ [$-3.214,\,-2.917$]
& $-6.39$ \\

$\langle \mathrm{CN}_{\mathrm{Li-O}}\rangle$
& $3.200 \pm 0.736$
& $4.263 \pm 0.651$
& $-1.062$ [$-1.275,\,-0.850$]
& $-1.53$ \\

$\langle q_{\mathrm{Li}}\rangle$
& $0.475 \pm 0.019$
& $0.456 \pm 0.016$
& $+0.018$ [$+0.013,\,+0.024$]
& $+1.04$ \\

$\langle q_{\mathrm{O,coord}}\rangle$
& $-0.346 \pm 0.036$
& $-0.309 \pm 0.033$
& $-0.037$ [$-0.048,\,-0.026$]
& $-1.07$ \\

$\langle E_{\mathrm{LES}}\rangle$ (eV)
& $-76.615 \pm 0.262$
& $-76.098 \pm 0.391$
& $-0.517$ [$-0.618,\,-0.415$]
& $-1.55$ \\

$\langle E_{\mathrm{short}}\rangle$ (eV)
& $-1524.887 \pm 0.707$
& $-1525.802 \pm 0.802$
& $+0.915$ [$+0.684,\,+1.150$]
& $+1.21$ \\

$\langle E_{\mathrm{total}}\rangle$ (eV)
& $-1601.502 \pm 0.713$
& $-1601.900 \pm 0.863$
& $+0.398$ [$+0.156,\,+0.641$]
& $+0.50$ \\

\bottomrule
\end{tabular}
\end{table}

%%%%%%%%%%%%%%%%%%%%%%%%%%%%%%%%%%%%%%%%%%%%%%%%%%%%%%%%%%%%%%%%%%%%%%%%%%%%%%%
\section{Discussion}
%%%%%%%%%%%%%%%%%%%%%%%%%%%%%%%%%%%%%%%%%%%%%%%%%%%%%%%%%%%%%%%%%%%%%%%%%%%%%%%

The central finding of this work is that the latent electrostatic variables
learned by MTP--LES retain reproducible information about ion-pairing and
solvation environments, even though the model is trained only on DFT energies
and forces~\cite{Zhong2025}. Associated and dissociated configurations of Li--PF$_6$ ion pairs in liquid electrolytes exhibit robust
ensemble-level differences in the latent Li response, the coordinating-oxygen
response, and the long-range LES energy, none of which were supplied as
explicit training targets. The latent electrostatic representation therefore functions as more than
an auxiliary numerical degree of freedom for constructing the total energy: it
encodes structure-resolved signatures of the environments represented
in the training data. This interpretation does not require identifying the
latent charges with any particular charge-partitioning scheme; instead, their
significance lies in reproducible relationships between the learned
variables and independently defined structural environments.

The energy decomposition provides a complementary perspective. The associated ensemble
exhibits an LES contribution approximately $0.5$~eV more favorable on average
than that of the dissociated ensemble, whereas its short-range contribution is
approximately $0.9$~eV higher. The two contributions therefore partially compensate in the total
model energy. This result illustrates how a model with separate short- and long-range channels
resolves energetic differences to the two ion-pair environments that are largely obscured in the total energy alone. Such a result should not, however, be interpreted as establishing a
thermodynamic preference for either ion-pair state. Since the two ensembles were sampled separately and the Li--P distances separating the two states are only sparsely covered, the observed offsets represent differences between sampled ensembles rather
than a potential of mean force. Establishing the latter will require
controlled sampling along an appropriate reaction coordinate together with proper treatment of the associated entropic contributions.

The coordinating-oxygen analysis indicates that the learned electrostatic
response cannot be reduced to a single nearest-neighbor geometric descriptor.
Although the latent oxygen response varies with O--Li separation, the
associated ensemble exhibits more negative latent oxygen charges across the
range of overlapping O--Li distances. The two ensembles also have similar
mean O--Li distances despite statistically distinct latent oxygen-charge distributions. Part of this difference reflects the composition of the Li$^+$ coordination
shell, which contains a larger share of EMC ether oxygens in the dissociated
ensemble. Together with the differences in Li--O coordination, the
mean first-shell oxygen response, and the distinct behavior of Li, P, F,
and coordinating oxygen atoms across the two environments, these results indicate that
the latent variables are sensitive to the broader ion-pair and solvation
environments rather than acting as mere proxies for a single local coordinate.

These observations complement previous work on explicit and
environment-dependent electrostatic MLIPs. Previous studies have shown that
LES can improve predictive accuracy relative to corresponding short-range
models in systems where long-range interactions are important. For example,
LES augmentation has improved the accuracy of CACE-based models for charged
and polar molecular systems, molten salts, and liquid
water~\cite{Cheng2025LES}, and LES has subsequently been demonstrated across
multiple MLIP architectures, including MACE, NequIP, CACE, and
CHGNet~\cite{Kim2025LESaug}. The predictive contribution of LES relative to
the corresponding short-range MTP architecture was also examined in our
previous MTP--LES study~\cite{AdekoyaOlowofela2026MTPLES}. In the context of
carbonate electrolytes, Chalykh \emph{et al.} further showed that
environment-dependent electrostatics can improve force accuracy and simulation
stability relative to a fixed-charge formulation, although these advantages
were not uniform across all downstream observables~\cite{Chalykh2026}. The present work addresses a complementary question: rather than reassessing predictive gains from electrostatic augmentation, we
ask what information is encoded in the learned electrostatic variables
themselves and how that information relates to the local ion-pair and solvation environment.

To assess whether the learned electrostatic signatures persist beyond the
original dataset, we applied the trained model without retraining to
the independent LiPF$_6$/EC/EMC dataset of Chalykh
\emph{et al.}~\cite{Chalykh2026}. This external dataset differs from the training data in salt
concentration (approximately 1~M versus 0.49~M), temperature and ensemble
(300~K in the NPT ensemble versus 330~K in the NVT ensemble), and exchange--correlation functional
(PBE-D3 versus RPBE-D3(BJ)). The model predicts external reference
forces with an RMSE of 104~meV\,\AA$^{-1}$ per Cartesian component, compared
with 56.6~meV\,\AA$^{-1}$ on our own held-out test set. The ensemble-mean latent
charges remain of comparable magnitude across the two datasets, including for configurations containing two ion pairs. In the external dataset, the latent Li charge ranges from 0.445 to 0.458,
compared with 0.456 to 0.475 in the held-out test set; the corresponding ranges for the coordinating oxygen response are $-0.348$ to $-0.341$ against $-0.346$ to
$-0.309$, respectively (Table~\ref{tab:statistics}). Using the contact criterion of Chalykh \emph{et al.}~\cite{Chalykh2026}, the
external latent Li charge again distinguishes contact and separated ion-pair environments, with a
mean difference of $+0.013$ (95\% CI [$+0.010$, $+0.016$]), compared with $+0.018$ in
our held-out test set (Table~\ref{tab:statistics}). The difference in Li--O
coordination between the two states is also similar in the two datasets, approximately $-1.06$.
Total energies are not directly compared because the two datasets 
use different DFT settings. The coordinating-oxygen contrast is smaller and
not statistically resolved in the external dataset. The LES-energy
contrast is likewise smaller, although differences in cell size limit direct
comparison of per-cell LES energies. Overall, these results provide evidence supporting model
transferability within the LiPF$_6$/EC/EMC electrolyte system, particularly
through the persistence of the latent Li-charge signature under conditions
beyond those represented in training.

Several limitations define the scope of this interpretation. The associated and
dissociated configurations were obtained from separate trajectories sampling
distinct ion-pairing environments. Consequently, the large difference in Li--P separation
follows from the construction of the dataset and serves as a structural
consistency check; the Li--P distances separating the two ensembles are sparsely sampled, and
apparent turning points in sparsely populated boundary bins should not be
assigned mechanistic significance. The latent charges are model-defined internal variables rather
than uniquely defined quantum-mechanical observables, and their absolute values
should not be equated with Bader, Hirshfeld, DDEC, or other conventional
partial charges~\cite{Bader1985,Hirshfeld1977,Manz2016}. Where absolute
electrical-response quantities are required, the latent-charge scale may
additionally require calibration associated with the electronic dielectric
response. Because total charge is enforced to remain neutral by construction, changes in the latent charges of different species are necessarily coupled, and therefore, their
correlated variation in these charges should not be interpreted as direct physical charge transfer between species.
Finally, although the external evaluation extends the conditions considered,
both datasets concern LiPF$_6$ in EC/EMC, and extending the analysis to other
salts, solvents, and interfacial environments is a natural next step.

Despite these limitations, our results demonstrate that a latent electrostatic
representation learned solely through energy and force supervision can retain
statistically robust and chemically relevant information about distinct ion-pairing and solvation
environments. Because these variables are evaluated during
potential evaluation, they can, in principle, be monitored during large-scale molecular
dynamics without requiring additional electronic-structure calculations or
post hoc charge-partitioning analysis. Internal electrostatic variables need not be assigned the status of unique physical partial charges to be
scientifically useful; instead, their value lies in their reproducible relationships with independent structural quantities.
From this perspective, MTP--LES provides not only an efficient way to
decompose learned short- and long-range energetic contributions but also a structure-resolved latent representation that can be interrogated for signatures of
the underlying electrolyte environment. More broadly, these findings highlight the potential of latent electrostatic representations to extend machine-learned interatomic potentials beyond energy and force prediction, providing new opportunities to investigate the interplay between local structure, electrostatic interactions, and chemical dynamics in complex materials systems.
%%%%%%%%%%%%%%%%%%%%%%%%%%%%%%%%%%%%%%%%%%%%%%%%%%%%%%%%%%%%%%%%%%%%%%%%%%%%%%%

%%%%%%%%%%%%%%%%%%%%%%%%%%%%%%%%%%%%%%%%%%%%%%%%%%%%%%%%%%%%%%%%%%%%%%%%%%%%%%%
\section{Methods}
%%%%%%%%%%%%%%%%%%%%%%%%%%%%%%%%%%%%%%%%%%%%%%%%%%%%%%%%%%%%%%%%%%%%%%%%%%%%%%%-------------

\subsection{Model development workflow}
\begin{figure}[H]
\centering
\includegraphics[width=\linewidth]{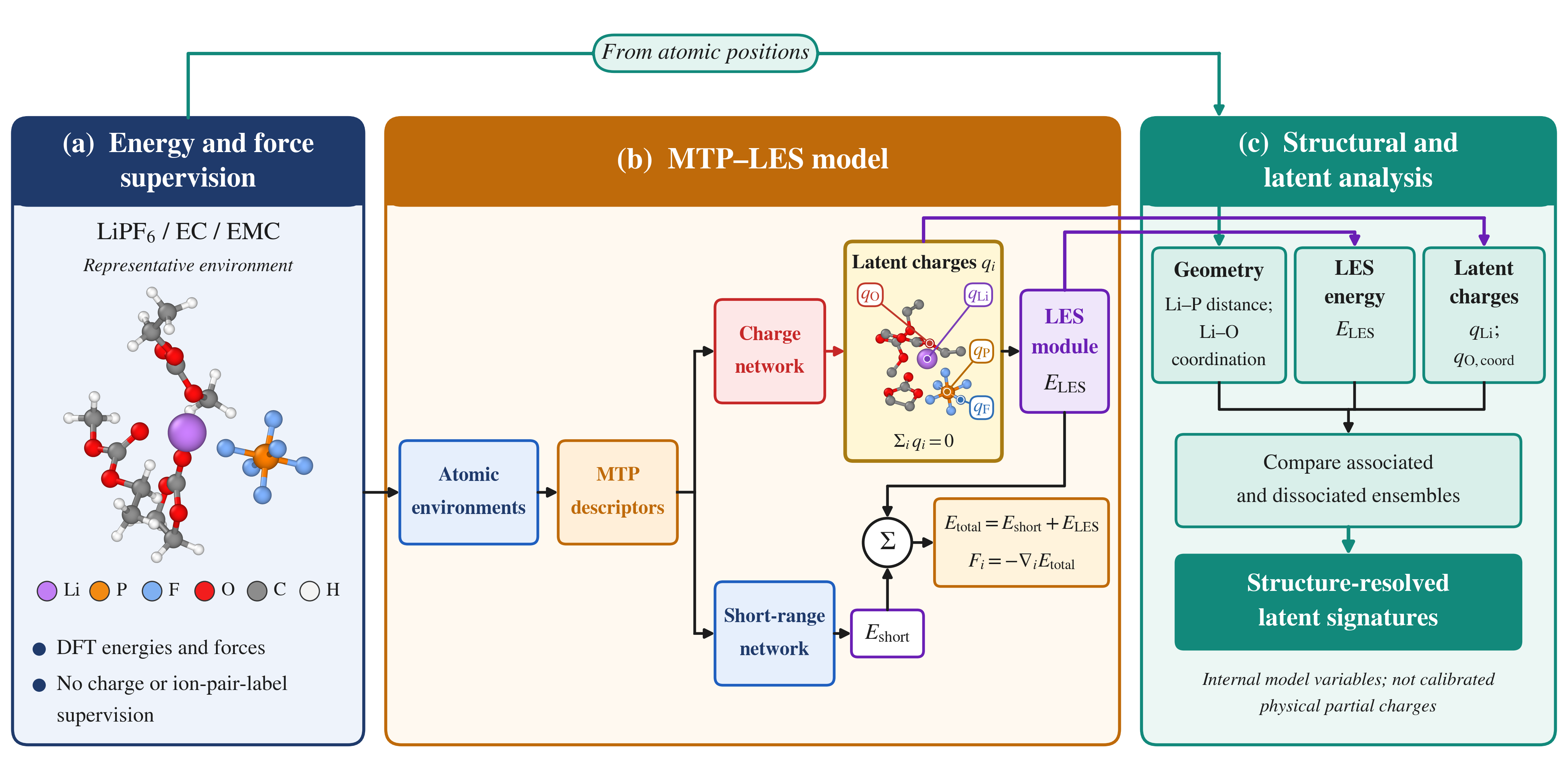}
\caption{\textbf{Model development and analysis workflow.} (a)~An MTP--LES
model is trained on DFT energies and forces of LiPF$_6$/EC/EMC configurations,
with no supervision on charges, dipoles, or ion-pair labels. (b)~The model
produces latent charges $q_i$ and the long-range energy $E_{\mathrm{LES}}$.
(c)~These emergent latent electrostatic variables are analyzed against
structural descriptors of ion association and solvation. Hydrogen atoms are
omitted in (b) for clarity.}
\label{fig:workflow}
\end{figure}

Figure~\ref{fig:workflow} summarizes the workflow of this study. The
DFT-labeled LiPF$_6$/EC/EMC configurations were split into training,
validation, and test sets of 1296, 144, and 160 configurations, each
containing both associated and dissociated environments. The MTP--LES model
was trained on total energies and atomic forces from the training set, the
validation set was used to monitor training and select the best checkpoint,
and the held-out test set was used only for the final evaluation of
predictive accuracy and the subsequent structural and latent-electrostatic
analyses.

% ------------------- 2) The model ------------------------------------

\subsection{The MTP--LES model}

We recently introduced MTP--LES, a charge-aware interatomic potential for molecular simulations of materials~\cite{AdekoyaOlowofela2026MTPLES}. The model expresses the total potential energy as the sum of short-range and long-range contributions,
\begin{equation}
E_{\mathrm{total}} \;=\; E_{\mathrm{short}} \;+\; E_{\mathrm{LES}},
\label{eq:etot}
\end{equation}
both computed from a common set of rotationally invariant Moment Tensor
Potential (MTP) descriptors $\{d_i\}$~\cite{Shapeev2016}. (These
contributions are denoted $E_{\mathrm{SR}}$ and $E_{\mathrm{LR}}$ in
Ref.~\cite{AdekoyaOlowofela2026MTPLES}.) Two independent
neural networks operate in parallel on these descriptors: a short-range
branch that predicts per-atom energies whose sum gives $E_{\mathrm{short}}$,
and a charge branch that predicts per-atom latent charges $\{q_i\}$ with a
learnable per-species bias. The raw charges are made exactly neutral for
each configuration by the differentiable projection
\begin{equation}
q_i \;=\; q_i^{\mathrm{raw}} - \frac{1}{N}\sum_{j=1}^{N} q_j^{\mathrm{raw}},
\label{eq:neutrality}
\end{equation}
which enforces $\sum_i q_i = 0$ without reference to any charge labels.

Following the Latent Ewald Summation
formulation~\cite{Cheng2025LES,Kim2025LESaug}, the long-range
energy is evaluated in reciprocal space as
\begin{equation}
E_{\mathrm{LES}} \;=\; \frac{1}{2\varepsilon_0 V}
\sum_{0<|\mathbf{k}|<k_c}
\frac{\exp\!\left(-\sigma^{2}k^{2}/2\right)}{k^{2}}\,
\bigl|S(\mathbf{k})\bigr|^{2}
\;-\; \frac{1}{4\pi\varepsilon_0}\,\frac{1}{\sqrt{2\pi}\,\sigma}\sum_i q_i^{2},
\qquad
S(\mathbf{k}) = \sum_i q_i\, e^{\mathrm{i}\mathbf{k}\cdot\mathbf{r}_i},
\label{eq:les}
\end{equation}
where $V$ is the cell volume, $\sigma$ is the Gaussian smearing width, and
the reciprocal-space cutoff $k_c$ is set by the grid spacing $\Delta l$. The
$\mathbf{k}=0$ term is omitted, corresponding to tinfoil boundary
conditions. The second term removes the Gaussian self-interaction of the
latent charges; because it is subtracted, $E_{\mathrm{LES}}$ can be negative.
No real-space pair term is included: the short-range physics is carried by
the local energy branch. Because the latent charges enter the model only through
Eq.~\eqref{eq:les}, they are never fitted to any charge-partitioning scheme;
they are internal degrees of freedom optimized end to end through the
energy and force losses.

Forces are obtained by automatic differentiation of $E_{\mathrm{total}}$
with respect to atomic positions, so the electrostatic contribution to the
forces is exactly consistent with Eq.~\eqref{eq:les}. The energy and force contributions to the training loss are
\begin{equation}
\mathcal{L} \;=\; w_E \mathcal{L}_E + w_F \mathcal{L}_F,
\label{eq:loss}
\end{equation}
where $\mathcal{L}_E$ and $\mathcal{L}_F$ are the mean-squared errors of
the energy per atom and the force components, with equal weights. A weak
charge-magnitude regularization ($\lambda_q = 10^{-3}$) is included as in
Ref.~\cite{AdekoyaOlowofela2026MTPLES}; it involves no reference charges or
electrostatic labels. Descriptor settings,
network architectures, and all training hyperparameters are listed in
Supplementary Table~S3; the full mathematical specification of the
descriptors is given in Ref.~\cite{AdekoyaOlowofela2026MTPLES}.

% ------------------- 3) Training -------------------------------------

\subsection{Training}

MTP--LES was trained using DFT total energies and atomic forces as the only
supervision targets, with the Adam optimizer for 100 epochs
(Supplementary Table~S3). Training was monitored on the validation set, and
the checkpoint with the lowest validation energy RMSE
(0.86~meV\,atom$^{-1}$) was retained for all analyses.

No reference atomic charges, electrostatic potentials, dipole moments, Born
effective charges, dielectric properties, Li--P distances, coordination
numbers, or ion-pair labels were included in the training objective.
Consequently, the latent electrostatic quantities analyzed in this work
emerge indirectly through optimization of the energy and force losses.

%%%%%%%%%%%%%%%%%%%%%%%%%%%%%%%%%%%%%%%%%%%%%%%%%%%%%%%%%%%%%%%%%%%%%%%%%%%%%%%
\subsection{Electrolyte dataset and AIMD simulations}
%%%%%%%%%%%%%%%%%%%%%%%%%%%%%%%%%%%%%%%%%%%%%%%%%%%%%%%%%%%%%%%%%%%%%%%%%%%%%%%

The dataset was generated from \emph{ab initio} molecular dynamics (AIMD)
simulations of LiPF$_6$ dissolved in a mixed ethylene carbonate (EC) and ethyl
methyl carbonate (EMC) solvent. Electronic-structure calculations were
performed using the revised Perdew--Burke--Ernzerhof (RPBE)
exchange--correlation functional~\cite{PhysRevB.59.7413} with Grimme's D3 dispersion
correction with the Becke-Johnson (BJ) damping~\cite{Grimme2011BJ}, as implemented in the Vienna \emph{Ab initio}
Simulation Package (VASP)~\cite{Kresse1996,Kresse1999}. Simulations were carried out under periodic boundary conditions using a cubic
cell of 15\,\AA{} $\times$ 15\,\AA{} $\times$ 15\,\AA{} including one Li$^+$/PF$_6^-$ ion pair
solvated by EC and EMC molecules, corresponding to 278 atoms.
A plane-wave cutoff of 450 \, eV and the Monkhorst-Pack k-point mesh grid of (2 × 2 × 1) were used.
The trajectories were propagated in the canonical ensemble (NVT) at
330 \,K using the Nose–Hoover thermostat and a timestep of
0.5 \,fs.

Separate AIMD trajectories were generated to sample contact ion-pair (CIP) and
solvent-separated ion-pair (SSIP) environments. The original trajectories
contained 40{,}010 associated (CIP) configurations and 43{,}593 dissociated
(SSIP) configurations. Because neighboring AIMD frames are strongly correlated, representative
configurations were selected independently from the two trajectories using
Farthest Point Sampling (FPS)~\cite{Ceriotti2013}. FPS was applied before data
partitioning to reduce temporal redundancy while retaining structurally diverse
configurations from each ensemble. A total of 800 configurations was retained
from each trajectory, giving a combined dataset of 1{,}600 configurations.
Applying FPS before partitioning reduced the likelihood that strongly
redundant AIMD configurations were distributed across different data splits. The selected configurations were randomly divided into training, validation, and test sets containing 1{,}296, 144, and 160 configurations, respectively.
The held-out test set contained 80 associated and 80 dissociated
configurations.

%%%%%%%%%%%%%%%%%%%%%%%%%%%%%%%%%%%%%%%%%%%%%%%%%%%%%%%%%%%%%%%%%%%%%%%%%%%%%%%
\subsection{Structural and latent-electrostatic analysis}
%%%%%%%%%%%%%%%%%%%%%%%%%%%%%%%%%%%%%%%%%%%%%%%%%%%%%%%%%%%%%%%%%%%%%%%%%%%%%%%

Configurations were assigned to the associated (CIP) or dissociated (SSIP)
ensemble according to their source AIMD trajectory. The Li--P separation was
used as a structural descriptor of ion association but was not supplied to the
model during training. Because the two source trajectories sample distinct
ion-pair environments, Li--P separation differs strongly between the ensembles
by construction and is therefore treated as a consistency check rather than an
independently discovered state discriminator.

Local Li$^+$ solvation was characterized using the Li--O coordination number,
defined as the number of solvent oxygen atoms within the first coordination
shell of Li$^+$. A cutoff of 2.65~\AA{} was used to define this shell, consistent
with the $\sim$2.65~\AA{} first-solvation radius reported from molecular
dynamics simulations of LiPF$_6$ in mixed carbonate
(EC/PC/DEC/EMC) electrolytes~\cite{Chen2023Solvation}.

For comparison, the pooled Li--O radial distribution function computed from
the full AIMD trajectories before subsampling shows a shallow minimum near
3.05~\AA{} (Fig.~\ref{fig:solvation}a). To test the sensitivity of the coordination
analysis to the chosen shell boundary, the calculation was repeated using
cutoffs from 2.65 to 3.20~\AA{}. The associated--dissociated differences retain
the same direction and remain statistically significant throughout this range
(Supplementary Table~S4). The nearest O--Li distance was also used as a local
geometric descriptor for the coordinating oxygen atoms.

Latent atomic charges and energy contributions were evaluated for every
configuration in the held-out test set. The latent charges of Li, P, F, and O
were analyzed as functions of Li--P separation, together with the long-range
LES energy, short-range energy contribution, and total model energy.

For the collective electrostatic analysis, the mean latent F charge was
computed within each configuration. The solvent response was characterized by
the mean latent charge of the first-shell oxygen atoms,

\begin{equation}
\left\langle q_{\mathrm{O}}\right\rangle_{\mathrm{shell}}
=
\frac{1}{N_{\mathrm{O,shell}}}
\sum_{i\in\mathrm{shell}} q_i ,
\end{equation}

where $N_{\mathrm{O,shell}}$ is the number of oxygen atoms in the first Li$^+$
coordination shell. Using the mean rather than the total shell charge separates
the per-oxygen latent response from changes in coordination number.

Latent charges are reported on the internal MTP--LES scale and are interpreted
as relative, structure-resolved descriptors rather than uniquely defined
physical partial charges. Because charge neutrality is enforced by construction,
correlated changes among atomic latent charges were not interpreted as direct
evidence of directional physical charge transfer between species.

%%%%%%%%%%%%%%%%%%%%%%%%%%%%%%%%%%%%%%%%%%%%%%%%%%%%%%%%%%%%%%%%%%%%%%%%%%%%%%%
\subsection{Binned analysis}
%%%%%%%%%%%%%%%%%%%%%%%%%%%%%%%%%%%%%%%%%%%%%%%%%%%%%%%%%%%%%%%%%%%%%%%%%%%%%%%

For visualization as a function of Li--P separation, the pooled Li--P range
was divided into eight equal-width distance bins. Statistics were then computed
separately for the associated and dissociated ensembles, resulting in the
state-specific disconnected curves shown in the figures. Bins containing fewer
than three configurations were excluded from the plotted curves; no populated
bins fell below this threshold.

The LES, short-range, and total energy contributions were analyzed using the
same state-specific binning procedure. For visualization of the energy
decomposition, energies were referenced to the mean of the dissociated
ensemble. These relative energies were used only to compare the sampled
ion-pair environments and were not interpreted as a potential of mean force or
thermodynamic free-energy profile.

%%%%%%%%%%%%%%%%%%%%%%%%%%%%%%%%%%%%%%%%%%%%%%%%%%%%%%%%%%%%%%%%%%%%%%%%%%%%%%%
%%%%%%%%%%%%%%%%%%%%%%%%%%%%%%%%%%%%%%%%%%%%%%%%%%%%%%%%%%%%%%%%%%%%%%%%%%%%%%%
\subsection{Coordinating-oxygen analysis}
%%%%%%%%%%%%%%%%%%%%%%%%%%%%%%%%%%%%%%%%%%%%%%%%%%%%%%%%%%%%%%%%%%%%%%%%%%%%%%%

To determine whether the latent solvent response could be explained solely by
local O--Li geometry, first-shell oxygen atoms were analyzed as a function of
their nearest O--Li separation. Because multiple coordinating oxygen atoms can
originate from the same configuration, oxygen-level observations were not
treated as statistically independent.

Uncertainty was estimated using a configuration-level clustered bootstrap with
10{,}000 resamples. Whole configurations were resampled with replacement, and
all qualifying coordinating oxygen atoms from each selected configuration were
retained together. This procedure was used for the quantitative ensemble
comparisons reported in the text and in Table~\ref{tab:statistics}. The binned
curves in Fig.~\ref{fig:solvation}c are shown with standard errors of the mean
(SEM). To account for differences in the sampled O--Li distance distributions, the
associated-minus-dissociated difference in coordinating-oxygen latent charge
was evaluated within O--Li distance bins and then combined across bins using
occupancy weights.

As a complementary configuration-level analysis, first-shell oxygen atoms with
O--Li distances greater than $2.0$~\AA{} were averaged within each
configuration before comparing the associated and dissociated ensembles. This
configuration-averaged quantity is the value reported in
Table~\ref{tab:statistics}. As a control for the neutrality projection, the mean latent charge of non-coordinating oxygen atoms was subtracted from that of the coordinating oxygen atoms within each configuration before comparing the two ensembles.

%%%%%%%%%%%%%%%%%%%%%%%%%%%%%%%%%%%%%%%%%%%%%%%%%%%%%%%%%%%%%%%%%%%%%%%%%%%%%%%
\subsection{Statistical analysis}
%%%%%%%%%%%%%%%%%%%%%%%%%%%%%%%%%%%%%%%%%%%%%%%%%%%%%%%%%%%%%%%%%%%%%%%%%%%%%%%

Structural descriptors and latent electrostatic quantities were compared
between the associated and dissociated ensembles using the held-out test set.
Means and standard deviations were computed independently for each ensemble,
and differences are reported throughout as associated minus dissociated.

Percentile confidence intervals were estimated using 10{,}000
configuration-level bootstrap resamples. Standardized
differences between ensembles were quantified using Cohen's
$d$~\cite{Cohen1988,Sawilowsky2009}, computed as the difference in ensemble
means normalized by the pooled standard deviation. For coordinating-oxygen
quantities, the clustered-bootstrap procedure described above was used to
preserve within-configuration dependence among oxygen atoms.
Configuration-level resampling does not account for residual correlation
between configurations drawn from the same trajectory.

\section*{Data and code availability}
The MTP--LES implementation, training and analysis scripts, and configuration
files will be made publicly available upon publication.

\section*{Acknowledgements}
The first-principles training data were generated by X. Q. at the Technical University of Denmark (DTU). This work was supported by the U.S. Department of Energy (DOE), Office of Energy Efficiency and Renewable Energy (EERE), through the High-Performance Computing for Energy Innovation (HPC4EI) program.
Work performed at the Center for Nanoscale Materials, a U.S. Department of Energy Office of Science User Facility, was
supported by the U.S. DOE, Office of Basic Energy Sciences, under Contract No. DE-AC02-06CH11357.
Computational support was provided by the National Energy Research Scientific Computing Center
(NERSC), a DOE Office of Science User Facility under Contract DE-AC02-05CH11231, as well as by the
Laboratory Computing Resource Center (LCRC) at Argonne National Laboratory.

\bibliography{bib}

\end{document}

\begin{center}

{\Large\bfseries Supporting Information\par}
\vspace{8pt}

{\large\bfseries
Interpreting Latent Electrostatics in Charge-Aware Machine
Learning Potentials for Battery Electrolytes\par}
\vspace{12pt}

{\normalsize
Abibat Adekoya-Olowofela$^{\dagger,\ddagger}$,
Sukriti Manna$^{\dagger,\ddagger}$,
Shola Folarin$^{\dagger,\ddagger}$,
Mikhail Trought\textsuperscript{\P},
Benjamin Emley\textsuperscript{\P},
Ann Straccia\textsuperscript{\P},
Avinesh Ojha\textsuperscript{*,\P},
Subramanian K.R.S. Sankaranarayanan$^{*,\dagger,\ddagger}$,
and Xueping Qin$^{*,\ddagger}$\par}
\vspace{8pt}

{\fontsize{9}{11}\selectfont\itshape

$^{\dagger}$Department of Mechanical and Industrial Engineering,
University of Illinois Chicago, Chicago, Illinois 60607,
United States\par

\vspace{3pt}

$^{\ddagger}$Center for Nanoscale Materials,
Argonne National Laboratory, Lemont, Illinois 60439,
United States\par

\vspace{3pt}

\textsuperscript{\P}Ford Motor Company, Dearborn,
Michigan 48124, United States\par}

\vspace{8pt}

{\small
$^{*}$Corresponding authors: aojha@ford.com;
skrssank@uic.edu; xqin@anl.gov\par}

\end{center}

\vspace{1em}

% ---------------- Table S1: prediction errors ----------------
\begin{table}[H]
\centering
\caption{\textbf{Prediction errors for the two ion-pair ensembles.} Energy
and force errors of MTP--LES relative to DFT on the held-out test set,
reported separately for the associated (CIP) and dissociated (SSIP)
configurations and for the combined set. Force RMSEs are per Cartesian
component and are also given by element.}
\label{tab:si_errors}
\footnotesize
\begin{tabular}{lccc}
\toprule
\textbf{Quantity} & \textbf{Associated (CIP)} & \textbf{Dissociated (SSIP)} & \textbf{All} \\
\midrule
Energy RMSE (meV\,atom$^{-1}$) & $1.07$ & $0.54$ & $0.85$ \\
Energy mean signed error (meV\,atom$^{-1}$) & $+0.98$ & $-0.29$ & $+0.34$ \\
Force RMSE (meV\,\AA$^{-1}$) & $56.3$ & $56.9$ & $56.6$ \\
\midrule
\quad Li & $67.2$ & $74.7$ & $71.1$ \\
\quad P  & $94.2$ & $73.8$ & $84.6$ \\
\quad F  & $36.5$ & $26.2$ & $31.8$ \\
\quad O  & $60.0$ & $60.8$ & $60.4$ \\
\quad C  & $74.1$ & $75.2$ & $74.6$ \\
\quad H  & $41.9$ & $42.6$ & $42.3$ \\
\bottomrule
\end{tabular}
\end{table}

% ---------------- Table S2: oxygen types ----------------
\begin{table}[H]
\centering
\caption{\textbf{First-shell oxygen atoms resolved by type.} Number of
Li--O contacts within the 2.65~\AA{} first-shell cutoff in the held-out test
set (80 configurations per ensemble) and mean latent charge for each oxygen
type. Oxygen atoms were classified as carbonyl or ether from their bonding
(one or two carbon neighbors) and assigned to EC or EMC from the composition
of their molecule. Charges were first averaged over the contacts of a given
type within each configuration and then over the configurations containing at
least one such contact. The values are descriptive means, reported without
uncertainty estimates. All first-shell oxygen atoms are included here, whereas
$\langle q_{\mathrm{O,coord}}\rangle$ in Table~1 of the main text is
restricted to O--Li distances greater than $2.0$~\AA{}; this accounts for the
difference between the last row and the values in Table~1. No EC ether oxygen
lies within the first shell in either ensemble.}
\label{tab:otype}
\footnotesize
\begin{tabular}{lcccc}
\toprule
 & \multicolumn{2}{c}{\textbf{Contacts}} & \multicolumn{2}{c}{\textbf{Mean latent charge}} \\
\cmidrule(lr){2-3}\cmidrule(lr){4-5}
\textbf{Oxygen type} & Associated & Dissociated & Associated & Dissociated \\
\midrule
EC carbonyl     &  80 &  44 & $-0.386$ & $-0.396$ \\
EMC carbonyl    & 120 & 160 & $-0.373$ & $-0.366$ \\
EMC ether       &  56 & 137 & $-0.276$ & $-0.267$ \\
\midrule
All carbonyl    & 200 & 204 & $-0.379$ & $-0.371$ \\
All first-shell & 256 & 341 & $-0.358$ & $-0.330$ \\
\bottomrule
\end{tabular}
\end{table}

% ---------------- Table S3: hyperparameters ----------------
\begin{table}[H]
\centering
\caption{\textbf{Architecture and training hyperparameters of the
MTP--LES model.} Notation follows Ref.~\cite{AdekoyaOlowofela2026MTPLES}.}
\label{tab:si_hyperparams}
\begin{tabular}{ll}
\toprule
\textbf{Parameter} & \textbf{Value} \\
\midrule
\multicolumn{2}{l}{\emph{Descriptors (invariant MTP)}} \\
Radial cutoff                                   & 4.5~\AA \\
Chebyshev radial polynomials $n_{\mathrm{rad}}$ & 12 \\
Radial channels $n_{\mu}$                       & 8 \\
Body order $\nu_{\max}$                         & 2 \\
Level cutoff $\ell_{\max}$                      & 24 \\
\midrule
\multicolumn{2}{l}{\emph{Model branches}} \\
Short-range branch (two residual blocks)        & (256, 128, 64) \\
Charge branch                                   & (64, 32), per-species bias \\
\midrule
\multicolumn{2}{l}{\emph{Latent Ewald summation}} \\
Gaussian smearing width $\sigma$                & 1.0~\AA \\
Reciprocal-space grid spacing $\Delta l$        & 2.0~\AA$^{-1}$ \\
Gaussian self-interaction term                  & subtracted \\
\midrule
\multicolumn{2}{l}{\emph{Training}} \\
Optimizer                                       & Adam \\
Initial learning rate                           & $3\times10^{-4}$, cosine-annealed to $10^{-5}$ \\
Batch size                                      & 4 \\
Weight decay                                    & $10^{-5}$ \\
Epochs                                          & 100 \\
Charge-magnitude regularization $\lambda_q$     & $10^{-3}$ \\
Train / validation / test configurations        & 1296 / 144 / 160 \\
\bottomrule
\end{tabular}
\end{table}

% ---------------- Table S4: cutoff sensitivity ----------------
\begin{table}[H]
\centering
\caption{\textbf{Sensitivity of coordination-based
results to the first-shell cutoff.} Associated-minus-dissociated differences
in the integer Li--O coordination number and the coordinating-oxygen latent
charge $\langle q_{\mathrm{O,coord}}\rangle$, recomputed on the held-out test
set for first-shell cutoffs $r_c = 2.65$--$3.20$~\AA. For
$\langle q_{\mathrm{O,coord}}\rangle$, only first-shell oxygens with
O--Li distances greater than 2.0~\AA{} were included, with values averaged
per configuration as in Table~1 of the main text. Confidence intervals were
obtained from 10{,}000 configuration-level bootstrap resamples. The
associated--dissociated differences retain the same direction, with 95\%
confidence intervals that exclude zero at all cutoffs.}
\label{tab:si_cutoff}
\footnotesize
\begin{tabular}{clcc}
\toprule
$r_c$ (\AA) & Quantity & $\Delta$ (assoc.\ $-$ dissoc.) [95\% CI] & Cohen's $d$ \\
\midrule
2.65 & CN$_{\mathrm{Li-O}}$ & $-1.062$ [$-1.275, -0.850$] & $-1.53$ \\
2.65 & $\langle q_{\mathrm{O,coord}}\rangle$ & $-0.037$ [$-0.048, -0.026$] & $-1.07$ \\
2.80 & CN$_{\mathrm{Li-O}}$ & $-1.050$ [$-1.250, -0.838$] & $-1.55$ \\
2.80 & $\langle q_{\mathrm{O,coord}}\rangle$ & $-0.041$ [$-0.051, -0.030$] & $-1.20$ \\
2.90 & CN$_{\mathrm{Li-O}}$ & $-1.013$ [$-1.213, -0.800$] & $-1.50$ \\
2.90 & $\langle q_{\mathrm{O,coord}}\rangle$ & $-0.041$ [$-0.051, -0.030$] & $-1.22$ \\
3.00 & CN$_{\mathrm{Li-O}}$ & $-0.838$ [$-1.038, -0.638$] & $-1.29$ \\
3.00 & $\langle q_{\mathrm{O,coord}}\rangle$ & $-0.035$ [$-0.045, -0.025$] & $-1.03$ \\
3.05 & CN$_{\mathrm{Li-O}}$ & $-0.788$ [$-0.987, -0.587$] & $-1.22$ \\
3.05 & $\langle q_{\mathrm{O,coord}}\rangle$ & $-0.032$ [$-0.042, -0.022$] & $-0.96$ \\
3.10 & CN$_{\mathrm{Li-O}}$ & $-0.700$ [$-0.900, -0.500$] & $-1.06$ \\
3.10 & $\langle q_{\mathrm{O,coord}}\rangle$ & $-0.031$ [$-0.041, -0.020$] & $-0.93$ \\
3.20 & CN$_{\mathrm{Li-O}}$ & $-0.562$ [$-0.775, -0.350$] & $-0.82$ \\
3.20 & $\langle q_{\mathrm{O,coord}}\rangle$ & $-0.030$ [$-0.039, -0.020$] & $-0.92$ \\
\bottomrule
\end{tabular}
\end{table}

\bibliographystyle{unsrt}
\bibliography{bib}